\documentclass[usenatbib,onecolumn,useAMS]{mn2e}
\usepackage{graphicx}    
\usepackage{latexsym}
\usepackage{dcolumn}
\usepackage{bm}
\input{epsf}  

\title[Precision era of the kinetic Sunyaev-Zeldovich effect]{Precision era of the kinetic Sunyaev-Zeldovich effect: simulations,
analytical models and observations and the power to constrain reionization}
\author[Pengjie Zhang, Ue-Li Pen \& Hy Trac]
{Pengjie Zhang\footnotemark, $^1$\thanks{E-mail:zhangpj@fnal.gov} Ue-Li Pen,
$^2$\thanks{E-mail:pen@cita.utoronto.ca} Hy Trac, $^2$,$^3$\thanks{E-mail:trac@cita.utoronto.ca}\\
$1$NASA/Fermilab Astrophysics Group,
Fermi National Accelerator Laboratory,
Box 500,
Batavia, IL 60510-050\\
$2$Canadian Institute for Theoretical Astrophysics, University of
Toronto, Toronto, Canada, M5S 3H8\\
$3$Department of Astronomy \& Astrophysics, University of
Toronto, Toronto, Canada, M5S 3H8}
\begin{document}      
\maketitle
\begin{abstract}
The kinetic SZ effect, which is the dominant CMB source
at arc-minute scales and $\nu \sim 217$ Ghz, probes the ionized gas peculiar momentum up to the epoch of  reionization and is a sensitive measure of the
reionization history. We ran high resolution self-similar and
$\Lambda$CDM hydro simulations and  
built an analytical model to study this effect.  Our model 
reproduces the $\Lambda$CDM simulation results to several percent
accuracy, passes various tests against self-similar simulations, and
shows  a wider range of applicability than
previous analytical models.  Our model in its continuous version is free of
simulation limitations such as finite simulation box and finite
resolution and  allows an 
accurate prediction of the kinetic SZ power spectrum $C_l$. For the WMAP
cosmology, we find $l^2C_l/(2\pi)\simeq 0.91 \times 10^{-12}
[(1+z_{\rm reion})/10]^{0.34}(l/5000)^{0.23-0.015(z_{\rm reion}-9)}$ for
the reionization redshift $6<z_{\rm reion}<20$ and $3000<l<9000$. The
corresponding temperature  fluctuation is several $\mu$K at these ranges. The
dependence of $C_l$ on the reionization history
allows an accurate measurement of the reionization epoch. For the Atacama
cosmology telescope experiment, $C_l$ can be measured with $\sim 1\%$
accuracy. $C_l$ scales as $(\Omega_b
h)^2 \sigma_8^{4\sim 6}$. Given cosmological
parameters, ACT would be able to  constrain   
$z_{\rm reion}$ with several percent accuracy. Some multi-reionization
scenarios degenerate in the primary CMB temperature and TE measurement can
be distinguished with $\sim 10 \sigma$ confidence. 

\end{abstract}
\begin{keywords}
Cosmic microwave background-theory-simulation: large scale
structure, intergalactic medium, reionization, cosmology
\end{keywords}

\section{Introduction}
Reionization is a critical phase in the evolution of the intergalactic medium
(IGM).
Recent constraints on the reionization epoch based on Lyman-$\alpha$
absorption and CMB appear to be in conflict.  
The WMAP measurement of the CMB temperature and polarization fluctuation
constrains the Thomson optical depth to the last
scattering surface $\tau\simeq 0.17 \pm 0.04 $. 
This large optical depth  suggests an early reionization at
$z\sim 17$ if the universe is re-ionized suddenly and remains completely
ionized  after that \citep{Kogut03,WMAP}.  This inference of the 
reionization redshift and optical depth are very robust: 
Thomson scattering is polarization dependent, and converts a temperature
quadrupole into a linear polarization.  The observed $\sim 2\mu K$ temperature
polarization cross correlation are just the product of the optical depths
and the local CMB quadrupole, $Q_{\rm rms,ps}=20\mu$K, 
which is known to high precision.  Given the optical depth, and the known
baryon content of the universe, one obtains a minimum redshift to
accumulate the sufficient total column density of electrons.

But, detection of a Gunn-Peterson
absorption trough in a SDSS quasar at redshift $z=6.28$ sets a constraint on
the reionization redshift at $z\sim6$ \citep{Becker01}.  
The quasar radiation is seen to be fully absorbed at a Lyman-$\alpha$ redshift
larger than 6.1.   This corresponds to a lower bound of neutral hydrogen
column density, corresponding to a neutral fraction greater than a percent.  At
face value, it would appear to be possible to have a small neutral fraction
while the majority of the electrons are ionized.  But if ionization
occurs through the UV radiation from discrete sources, this radiation
must be able to penetrate the IGM to ionize it.  It is generally not
possible for the radiation to pass through an opaque medium to ionize
material behind it.  One would generically expect the universe to be
either mostly neutral, or fully transparent with neutral fractions less than
$10^{-3}$.  
Furthermore, in order to keep the IGM temperature $T\sim 2\times 10^4$ K at $z\sim2$-$4$
required  by the Lyman-$\alpha$ forest observation,   a late time 
reionization must occur after $z=10$ \citep{Hui03}.  

Exotic possibilities exist to reconcile a sudden reionization with both
WMAP polarization and SDSS quasar data.  The ionization could arise from
decaying dark matter, which would be uniformly distributed.  Or it could
have ionized multiple times, and recombined in-between.
All these arguments lead to more complicated reionization scenarios.

To make observational progress on this conundrum, more information about
the fraction and dynamics of free electrons at $z\sim 10$ is needed.
The primary CMB temperature measurement depends on the reionization
history only through $\tau$ and thus has an intrinsic degeneracy. A precision  CMB
E-mode polarization measurement is in principle able to break this degeneracy
\citep{Hu03}. But the weak polarization signal makes such measurement 
unfeasible in
the near future. Due to the huge Lyman-$\alpha$ optical depth which must be
present in any multi-reionization scenarios,  the Lyman-$\alpha$ forest can
only detect the last 
reionization and makes it an indirect probe of the reionization
history.  The $21$ cm emission and  absorption backgrounds in the dark age
\citep{Tozzi00,Iliev02,Iliev03,Chen03} would constrain the upper limit of the
first reionization redshift. The contamination by free-free
emission of ionizing halos  \citep{Oh03} may be distinguished by their spectral
features. Such a measurement may be possible with a square kilometer array, but
its construction is still far in the future.

Free electrons resulting
from reionization also
scatter off CMB photons through Compton scattering and result
in the kinetic Sunyaev Zel'dovich 
(SZ) effect. 
Compton scattering keeps
information from high redshifts because it does not depend on redshift and is not affected by
distance or the expansion of the universe.  Since free electrons at high
redshifts have higher number density, their contribution to the kinetic SZ
effect is larger.  So the kinetic SZ effect is a sensitive measure of
the reionization process at high redshifts. Next
generation CMB experiments such as \citet{ACT} are making its precision
measurement feasible in the foreseeable future.  Once we have an accurate model
of this effect, the reionization history will be precisely constrained. It is
the goal of this paper to build such a model and to quantify the power of the
kinetic SZ effect to constrain the reionization history.

Unlike most radiative processes,
the  physics of the kinetic SZ effect is very simple.
In contrast to models of the Lyman-$\alpha$ forest,
the ionized gas peculiar momentum does not require an accurate understanding
of the gas state such as metalicity, temperature  and ionization 
equilibrium. 
In the linear regime, the exact model
is known as the Vishniac effect \citep{Vishniac87}. 
The non-linear regime is intrinsically more difficult and
theoretical works  in the literature have yet to converge quantitatively
\citep{Hu00,daSilva01a,daSilva01b,Gnedin01,Springel01,Valageas01,
Ma02,Zhang02}.

Hydrodynamical simulations are best equipped to capture the nonlinear IGM
physics.  Large box sizes are required to probe large-scale
power while high resolution is needed to resolve small-scale, nonlinear
structures. Most simulations to date have sacrificed one for the other.  But
the kinetic SZ effect requires the capturing of both large and small scale
structures and puts a strong requirement on the simulation
power. Limited computational resources introduce numerical artifacts 
which must be quantified. In
contrast, analytical models can be constructed to span all relevant scales
continuously, but they are often {\it ad hoc} procedures and must be tested and
calibrated against simulations. In this paper,  we run Eulerian hydrodynamical
simulations with box sizes of 50 and 100 Mpc/$h$ to model the evolution of the
IGM and measure the gas momentum for various redshifts.  Numerical limitations
such as finite simulation box size, finite resolution, and non-linearity are
studied by running different resolution simulations, ranging from $256^3$ up to
$1024^3$ fluid elements.  Self-similar simulations are also used to quantify
the numerical limitations. We then build an analytical model covering both
linear and nonlinear regimes, motivated by the Vishniac effect and
simulations. Our model is free of numerical artifacts and captures the main
effect of the non-linearity at the same time.  

In $\S$\ref{sec:formalism} we review the formalism for the KSZ effect.  In
$\S$\ref{sec:simulation} we describe the numerical simulations, including their
results and their limitations. Motivated by the gas hydrodynamics and
simulations, we build an analytical model in $\S$\ref{sec:theory}.  This model
is tested against simulations in \S \ref{sec:comparison}. In
$\S$\ref{sec:cl} we calculate the KSZ power
spectrum using our analytical model and in $\S$\ref{sec:ACT} we discuss the
potential of the KSZ effect to constrain the reionization history.

\section{The Kinetic SZ Effect: formalism}
\label{sec:formalism}
The kinetic SZ effect directly probes the ionized electron peculiar momentum
${\bf 
p}\equiv(1+\delta_e) {\bf v}$ through Compton scattering. The resulting CMB
temperature fluctuation is given by 
\citet{Sunyaev80}:
\begin{equation}
\Theta(\hat{n})\equiv \frac{\Delta T}{T_{\rm CMB}}=\int \chi_e \bar{n}_e \sigma_T \frac{{\bf p}\cdot
\hat{n}}{c}  \exp[-\tau(z)] a{\rm d}x=\int \frac{{\bf p}\cdot
\hat{n}}{c} \exp[-\tau(z)] {\rm d}\tau .
\label{eqn:ksz}
\end{equation}
Here, $\bar{n}_e$ is the mean electron number
density and  $\chi_e$ is the ionization fraction. $\sigma_T$ is
the Thomson cross section. $\hat{n}$ is the direction on the sky and
$x$ is the comoving distance. $\tau(z)=\int_0^z  \chi_e \bar{n}_e \sigma_T a
{\rm d}x$ is the Thompson optical depth.  ${\bf p}$ can be decomposed into a
curl-free or gradient part ${\bf p}_E$  satisfying $\nabla \times {\bf p}_E=0$
and a divergence-free or curl part ${\bf p}_B$ satisfying $\nabla
\cdot{\bf p}_B=0$.  The gradient term cancels out when integrating along the
line of sight and has no
contribution to the kinetic SZ effect. 
Thus the only part which contributes to the integral (\ref{eqn:ksz}) is ${\bf
p}_B$. We define the correlation functions  $\xi_{B,E}(r)\equiv \langle {\bf
  p}_{B,E}({\bf x})\cdot 
{\bf p}_{B,E}({\bf x+r})\rangle$  and the corresponding power spectra
$P^2_{B,E}(k)\equiv \langle |{\bf p}_{B,E}({\bf k})|^2 \rangle$. Throughout
this paper, we use $\Delta^2_{B}(k,z)\equiv k^3 P^2_{B}(k)/(2\pi^2)$ as the
gas momentum curl part power spectrum instead of $p^2_{B}(k)$. 
$\Delta^2_B(k,z)$ is a detailed description of the IGM state, whose 2D
projection is the kinetic SZ power spectrum $C_l$ (Eq. \ref{eqn:cl}). 

Adopting Limber's approximation  and the fact that
$\langle 
|{\textbf p}_{B}({\textbf k})\cdot \hat{n}|^2\rangle=\frac{1}{2} P^2_B(k)$, we
obtain the kinetic SZ angular correlation function: 
\begin{eqnarray}
w(\theta)&\simeq&
\left(\frac{\sigma_T}{c}\right)^2 \cos  \theta \int_0^{x_{\rm{re}}} dx
[a \bar{n}_e \bar{\chi}_e]^2 \exp[-2 \tau(z)] \int_{-\infty}^{\infty}
\frac{1}{2}\xi_B(\sqrt{x^2 \theta^2+y^2}) dy \\ \nonumber
&\simeq& \left(\frac{\sigma_T}{c}\right)^2  \int_0^{x_{\rm{re}}} dx [a \bar{n}_e\bar{\chi}_e]^2
\exp[-2 \tau(z)] \int_{-\infty}^{\infty}
\frac{1}{2} \xi_B(\sqrt{x^2 \theta^2+y^2}) dy.
\end{eqnarray}
Here, $x_{\rm{re}}$ is the comoving distance to the reionization
epoch.  We have set $\chi_e=\bar{\chi}_e$, namely, omitted the patchy
reionization effect\footnote{We will discuss the patchy reionization
in \S \ref{sec:ACT}.} following  the suggestion of both simulations
\citep{Gnedin01} and theory \citep{Valageas01}. 
The last approximation introduces an error less than $0.2 \%$
for $\theta \leq 1^0$.
Thus, the usual expression of the Limber's equation \citep{Peacock99}
still holds:
\begin{equation}
\label{eqn:cl}
C_l= \frac{16 \pi^2}{(2l+1)^3}[\frac{\bar{n}_e(0)\sigma_T}{c}]^2 
 \times \int_o^{z_{\rm{re}}}
(1+z)^4 \bar{\chi}_e^2 \frac{1}{2} \Delta^2_B(k,z)|_{k=l/x}
   \exp[-2 \tau(z)] x(z) \frac{dx(z)}{dz} dz. 
\end{equation}

${\bf p}_B$ is zero in first order linear perturbation theory. Thus, the
kinetic SZ effect is an intrinsically
non-linear process which must be investigated by either hydro simulations or
non-linear theories such as high order perturbation theory. After the 
pioneering work of the Vishniac effect \citep{Vishniac87}, 
which is a robust prediction of the
kinetic SZ effect in the linear regime, various authors \citep{Hu00,daSilva01a,daSilva01b,Gnedin01,Springel01,Valageas01, 
Ma02,Zhang02} have studied the kinetic SZ effect in the non-linear regime with analytical
models and simulations, but  there are still some unclear key issues. Three key issues we will address in this paper are (1) How
well do simulations capture $\Delta^2_B$ and thus the large scale power
spectrum of the kinetic SZ effect at large and small scales?  $\Delta^2_B$  has contributions from both large
and small scales. Simulations have cutoffs at both scales due to
the finite simulation box size and resolution, respectively. These cutoffs
affect 
simulation results and need to be quantified. Then, how to quantify and
overcome these effects to produce reliable results? (2) What is the effect
of the non-linearity to $\Delta^2_B$?  How to model this effect?  (3) How
observable is the effect, and to what accuracy can it be used to model
measured reionization by upcoming experiments? We will address these issues using simulations (\S \ref{sec:simulation}) and
our analytical model (\S \ref{sec:theory}).

\section{Hydrodynamic simulations}
\label{sec:simulation}
We ran cosmological hydrodynamical simulations using a new Eulerian 
cosmological hydro code  \citep{trac03a,trac03b}.  This Eulerian code
(hereafter TP) is based on the finite-volume, flux-conservative total variation
diminishing (TVD) scheme that provides high-order accuracy and high-resolution
capturing of shocks.  The hydrodynamics of the gas is
simulated by solving the Euler system of conservation equations for mass,
momentum, and energy on a fixed Cartesian grid.  The gravitational evolution of
the dark matter is simulated using a cloud-in-cell particle-mesh (PM) scheme
\citep{hockney88}.

The robustness of  the TP code has been tested by comparing the evolution of
the dark matter and gas density power spectra from the simulations with the
fitting formula of \citet{Smith02}.  We also performed 
a code comparison by running
the same initial conditions using the MMH code \citep{Pen98}, which
combines the 
shock capturing abilities of Eulerian schemes with the high dynamic range in
density achieved by Lagrangian schemes.  Power spectra are computed using FFTs.
We find good agreement at all relevant scales and redshifts for both
comparisons.

The KSZ effect is a non-local problem and has contributions from both
large-scale and small-scale power.  This makes it a challenging problem to
model numerically because high resolution is needed at all scales.  Eulerian
schemes provide high mass resolution at all scales, unlike Lagrangian schemes
based on SPH which achieve high dynamic range in high-density regions but do
poorly in low-density regions.  Eulerian schemes are ideal for simulating the
evolution of the IGM to model the thermal and kinetic SZ effects and
the Lyman alpha  
forest, because of their high speed, superior mass resolution,
shock-capturing abilities.  Furthermore, Eulerian algorithms
are computationally very fast and memory friendly, allowing one to optimally
use available computational resources.

We ran a total of six simulations with the best fit WMAP-alone
cosmology $\Omega_m=0.268$, 
$\Omega_\Lambda=0.752$, $\Omega_b=0.044$, $h=0.71$, and $\sigma_8=0.84$ \citep{WMAP}. Two
box sizes of 50 and 100 Mpc/$h$ are used and for each we perform the
simulations on fixed grids with $256^3$, $512^3$, and $1024^3$ cells.  The
ratio of dark matter particles to fluid elements is 1:8 in all six $\Lambda$CDM
simulations.  In the highest resolution simulation, we achieve a spatial
resolution of $\Delta x\simeq50$ kpc/$h$ and a dark matter particle mass
resolution of $\Delta m\simeq7\times10^7\ M_\odot$.  The initial conditions are
generated by sampling from an initial power spectrum computed using CMBFAST
\citep{CMBFAST}.  The simulations are started at a redshift of $z=100$ and
evolved down to $z=1$, with data outputs at $z=$ 20, 10, 6, 3, and 1.  The KSZ
effect, for angular scales $l\ga1000$, has contributions from electron
scattering primarily from the $z\ga1$ IGM and therefore, we do not push the
simulations below $z=1$ for the purposes of this paper. Furthermore, the final
prediction of the kinetic SZ effect is calculated using our analytical model,
instead of using these simulations directly, where the contribution from $z<1$
IGM is included.  The highest resolution
simulation takes approximately 700 time-steps to evolve from $z=100$ down to
$z=1$.  On a GS320 Compaq Alpha server with 32 cpus and total theoretical peak
speed of 32 Gflops, the run takes approximately 2 days.

\begin{figure}
\epsfxsize=10cm
\epsffile{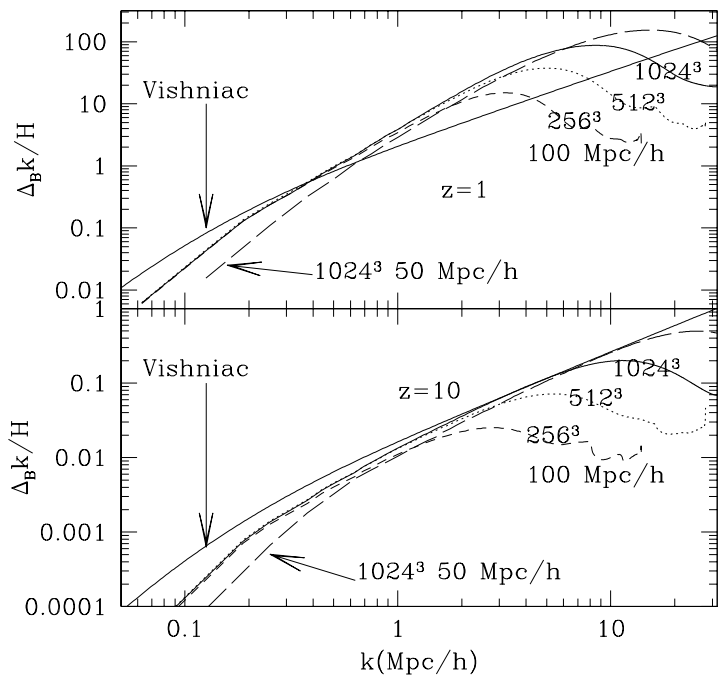}
\caption{The simulated $\Delta_B(k) k/H(z)$ at $z=1$ and $z=10$ for four of our
simulations. We are plotting the standard deviation $\Delta_B$ and not the
variance $\Delta_B^2$ which is usually plotted. Comparing to the
Vishniac prediction, which is accurate at linear 
scales, simulations lose power at these scales.  The $50$ Mpc/h simulation loses
more power than the $100$ Mpc/h simulations and suggests the effect of the
finite simulation box size.   At small scales, the simulation resolution effect is shown in the large $k$ behavior of
the three $100$ Mpc/h simulations.  Simulated $\Delta_B(k)$ begins to lose
power at less than $10$-$20$ grid separation.   $\Delta_B(k) k/H(z)\sim 1$
signals the linear-nonlinear transition region. Non-linear effects increase
$\Delta^2_B(k)$ relative to Vishniac theory for the WMAP cosmology we adopted. 
We also notice that the sign of the correction depends on the slope of
the power spectrum. 
\label{fig:simulation}}
\end{figure}

The peculiar gas momentum is decomposed into the rotational (B) and
irrotational 
terms (E) in Fourier space.  In Fig. \ref{fig:simulation} we plot the
simulated dimensionless $\Delta_B(k,z)k/H(z)$ at $z=1$ and $10$ and compare
them to the linear Vishniac effect, which is a robust prediction in the linear
regime (\S \ref{sec:theory}).   $H(z)$ is
the Hubble parameter. $\Delta_B(k,z)k/H(z)$ is the velocity of a wave
expressed in units of wave lengths per expansion time.
It reaches $\sim1$ when the overdensity reaches the mildly
nonlinear value $\delta\sim1$.  We note that our convention differs from 
the literature that we plot the root-mean-square standard deviation instead
of a variance.

Three key conclusions can be drawn from our simulations. (1) Finite 
simulation box
size causes $\Delta^2_B(k,z)$ to lose power at large scales.
At redshift $z=10$, where linear theory is a good approximation, we would
expect the simulation results to agree well
with the Vishniac prediction. But it turns out that they agree only at a very
limited $k$ range. More counter-intuitively, the discrepancy diverges toward
large scales. The simulated $\Delta^2_B(k,z)$ loses power toward large
scales. 50 Mpc/$h$ simulations lose more power at large
scales than 100 Mpc/$h$ simulations at the same scales. This suggests the effect of the finite
simulation box size. Simulations cut off density fluctuation modes larger
than half simulation box size and thus affect $\Delta^2_B$ through its non-local
dependence on the density and velocity fields. (2) Finite simulation resolution
causes $\Delta^2_B(k,z)$ to lose power at small scales (less than $\sim10$-$20$ grid separation). $\Delta^2_B(k,z)$ truncates
at small scales and the truncating scale $k_{\rm trunc}$ keeps increasing with
respect to the simulation resolution. Gas pressure is expected to
decrease power 
at very small scales. If this truncation is caused by the gas pressure, we
would expect  $k_{\rm trunc}$ to decrease with the simulation resolution  since
higher resolution results in higher gas pressure.  Since we observe
opposite behavior, we believe that this
truncation is caused by the simulation resolution, which predicts that $k_{\rm
trunc}$ to be roughly proportional to the resolution, as apparent in
figure. \ref{fig:simulation}.  (3) The non-linearity increases
$\Delta^2_B(k,z)$ with respect to the linear prediction.  This is visible
in the top panel of figure. \ref{fig:simulation}.

We further investigate these issues by self similar simulations. We ran two
self similar $512^3$ simulations with $n=-1$ and $n=-2$,
respectively. $\Omega_m=1.0$. $\Omega_b=0.164$ is adopted such that
$\Omega_b/\Omega_m$ is the same as the WMAP cosmology . The effects of finite simulation box size and
finite simulation resolution scale as expected. The non-linear effect turns
out to be more subtle. For $n=-1$, simulated $\Delta^2_B$ is smaller than the
Vishniac prediction in the non-linear regime. The non-linearity decreases $\Delta^2_B$. For $n=-2$,  simulated $\Delta^2_B$ is larger than the
Vishniac prediction. Thus the non-linearity  increases $\Delta^2_B$. To quantify simulation
artifacts and the non-linearity effect, we build an analytical model in \S
\ref{sec:theory}. Therefore we postpone the discussion and explanation of the
self similar simulations until in \S \ref{sec:comparison}.

\section{The analytical model} 
\label{sec:theory}
In this section, we will extend the Vishniac result to the grid space to model
 the simulation effect (\S \ref{subsec:linear}) and to the non-linear regime to
 model the non-linearity effect (\S \ref{subsec:nonlinear}).

\begin{figure}
\epsfxsize=10cm
\epsffile{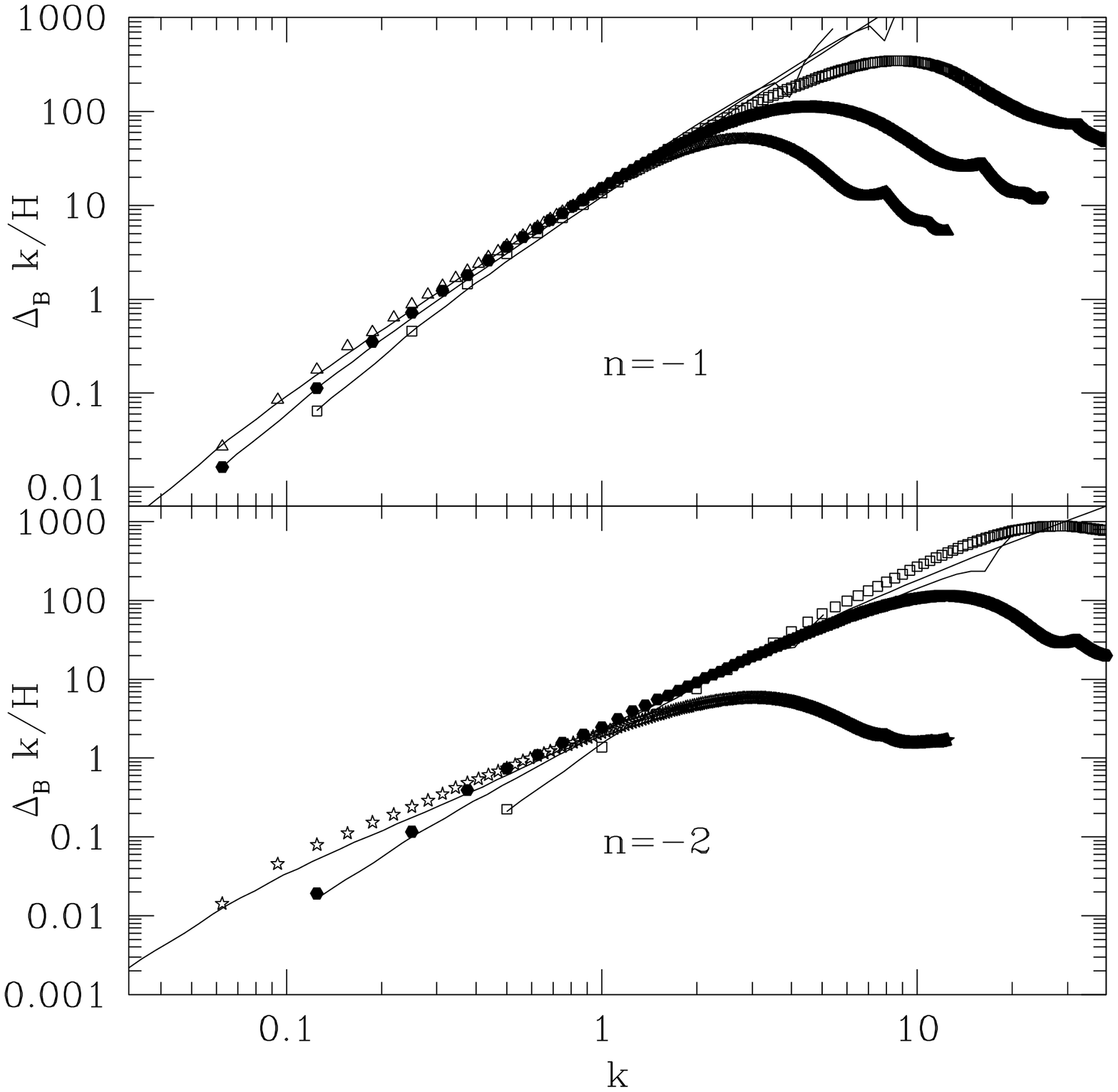}
\caption{The comparison between our model and $512^3$ grid elements self similar
simulations. Solid lines are the grid prediction of our model. The
effect of box size and resolution is clearly shown in this figure. As
a reminder, we are plotting the standard deviation $\Delta_B$ and not the
variance $\Delta_B^2$ which is usually plotted. \label{fig:self}}
\end{figure}

\subsection{Linear regime}
\label{subsec:linear}
In the linear regime, the motion of IGM is determined by the gravitational
 potential $\phi$. Thus ${\bf v} \propto \nabla \phi$ and is curl-free. Since $\nabla^2 \phi \propto \delta$,  ${\bf
v}({\bf k})\propto \delta_k {\bf k}/k^2$. From this relation, $\Delta^2_B$ in
 the linear regime  was  originally calculated by \citet{Vishniac87}.  With the correction of the gas window function $W_g$, we obtain
\begin{equation}
\label{eqn:conpb}
\Delta^2_B(k)=\frac{k^3}{2\pi^2}\frac{1}{2}
\frac{(a\dot{D}/D)^2}{(2\pi)^3}\int_0^{\infty}
P_L(k_2,z)P_L(|\textbf{k}-\textbf{k}_2|,z) 
\left[ W_g(|\textbf{k}-\textbf{k}_2|)\beta(\textbf{k,k}_2)+W_g(k_2)\beta(\textbf{k},\textbf{k}-\textbf{k}_2)
\right]^2 {\rm d}^3 k_2 . 
\end{equation}
$D$ is the linear growth factor of the dark matter over-density, 
$\dot{D}$ is its time derivative and $a$ is the scale factor of the
universe. $P_L(k,z)$ is the linear dark matter power spectrum. Gas pressure originated from various non-gravitational
heating processes smooths out the gas density at small scales. This effect
can be modeled by a window
function $W_g(k)$ such that $P_g(k)=W_g^2(k) 
P_{\rm DM}(k)$, which can be choose as a Gaussian window function (see e.g. \citet{Hui97,Gnedin00,Gnedin02}). 
In an adiabatically
evolving universe, gas traces dark matter down to very small scales and $W_g$ would
be effectively unity. One can model the simulation resolution by
introducing  an
effective grid smoothing window $W_g$. This procedure works well but the modeling of $W_g$ is arbitrary.
For simplicity, we omit this gas window effect. 
Since the gas peculiar
velocity ${\bf v_g}$ is mainly determined by the dark matter gravitational potential field, we have assumed that ${\bf v_g}$ traces the dark
matter peculiar velocity at all scales. The kernel $\beta({\bf k,k_2})={\bf(
  k_2-\frac{k(k\cdot k_2)}{k^2})/k_2^2}$. This form originates from the
gradient operators in ${\bf
v}\propto \nabla \phi$ and $\nabla^2 \phi \propto
\delta$. Eq. (\ref{eqn:conpb}) 
explicitly preserves the symmetry between ${\bf k}_1\equiv{\bf k}-{\bf k}_2$ and
${\bf k}_2$.

Eq. (\ref{eqn:conpb}) is exact in the
linear regime and an ideal simulation should be able to reproduce these
results. But we notice from this equation that $\Delta^2_B$ has contributions
from both large  and small scales. For a real simulation, the
simulation box size and resolution cut off contributions from these
scales  and the simulation result may deviate from Eq. (\ref{eqn:conpb}).  To
address this issue and to estimate the limitation of simulations, we
calculate $\Delta^2_B$ on discrete grids.  For a
simulation with 
box size $L$, resolution $N$ and the periodic boundary condition,  the Fourier
component of the gradient operator 
$\nabla$ is $\tilde{\nabla}=-i\sin({\bf k}\Delta L)/\Delta
L\equiv-i\left\{\sin(k_x\Delta 
L)/\Delta L,\sin(k_y\Delta L)/\Delta L,\sin(k_z\Delta L)/\Delta
L\right\}$. Here, $\Delta L\equiv L/N$. Comparing to
$\tilde{\nabla}=-i{\bf k}$ in the continuous 
case,  we obtain the grid version of $\Delta^2_B$:
\begin{equation}
\label{eqn:gridpb}
\Delta^2_B(k)=\frac{k^3}{2\pi^2}\frac{1}{2}
(a\dot{D}/D)^2 L^3 \sum_{\textbf{k}_2} P_L(k_2,z)P_L(|(\textbf{k}-\textbf{k}_2)_p|,z)
\left[W_g(|(\textbf{k}-\textbf{k}_2)_p|)\beta_G(\textbf{k,k}_2)+W_g(k_2)\beta(\textbf{k},(\textbf{k}-\textbf{k}_2)_p)
\right]^2.
\end{equation}
Here, $\beta_G({\bf k,k_2})=\beta({\bf k}\rightarrow \sin({\bf
k}\Delta L)/\Delta L,{\bf k_2}\rightarrow\sin({\bf k_2}\Delta
L)/\Delta L)$, $k_{x,y,z},k_{2x,y,z}=2\pi/L\times(-N/2,-N/2-1,\ldots,
N/2-1)$. $(\textbf{k}-\textbf{k}_2)_p$ means 
$\textbf{k}-\textbf{k}_2$ under the 
periodic boundary condition. Namely, $\textbf{k}-\textbf{k}_2$ needs to be
converted to the same range as 
$\textbf{k}_2$ by the periodic boundary condition. This grid version represents
simulation discretization effect at the first order approximation.

\subsection{Non-linear regime}
\label{subsec:nonlinear}
In the non-linear regime, the perturbation theory does not hold
and Eq. (\ref{eqn:conpb}) \& (\ref{eqn:gridpb}) break 
down. \citet{Hu00,Ma02} argue
that the velocity field is less non-linear than 
the density field and one may substitute $P_L(k,z)$ in Eq. (\ref{eqn:conpb}) 
introduced by the density field with the corresponding non-linear density power
spectrum $P_{NL}(k,z)$  and keep the other $P_L$ introduced by the velocity
field unchanged. But the generation of curl in the velocity field by shell
crossing in the non-linear regime weakens this argument.  In the presence of curl, $\nabla \times {\bf p}_B=\nabla \delta\times {\bf
v}_E+\nabla \delta \times {\bf v}_B+ (1+\delta) \nabla\times {\bf v}_B$. The
first term in the right side of this equation is well described by the linear Vishniac
effect and may be well described by the method of \citet{Hu00,Ma02} in
the non-linear regime. But the  remaining terms 
bring non-negligible contribution in the non-linear regimes.  In our
simulations, we find that  
in the highly non-linear regime, velocity field reaches equi-partition and
$\Delta^2_{{\bf v}_B}=2 \Delta^2_{{\bf v}_E}$. This suggests that the
contribution from ${\bf v}_B$ may be dominant in the non-linear regime. Thus,
the method of \citet{Hu00,Ma02} underestimates $\Delta^2_B$. The  generation of
${\bf v}_B$ through shell crossing involves complicated gas hydrodynamics
and makes the 
analytical understanding of ${\bf v}_B$ intractable. Its statistics
in principle can only be probed by simulations. But since the
generation of ${\bf v}_B$ is caused by the non-linearity, to model its effect, one may
phenomenologically substitute the linear density power spectrum in
Eq. (\ref{eqn:conpb}) introduced by the velocity field with the corresponding
non-linear density power spectrum. This procedure at least qualitatively
captures the effect of ${\bf v}_B$ generation. Thus we obtain the  final
continuous form  of $\Delta^2_B(k,z)$:
\begin{equation}
\label{eqn:finalconpb}
\Delta^2_B(k,z)=\frac{k^3}{2\pi^2}\frac{1}{2}
\frac{(a\dot{D}/D)^2}{(2\pi)^3}\int_0^{\infty}
P_{NL}(k_2,z)P_{NL}(|\textbf{k}-\textbf{k}_2|,z) 
\left[W_g(|\textbf{k}-\textbf{k}_2|)\beta(\textbf{k,k}_2)+W_g(k_2)\beta(\textbf{k},\textbf{k}-\textbf{k}_2)\right]^2{\rm
  d}^3 k_2
\end{equation}
and the corresponding grid form 
\begin{equation}
\label{eqn:finalgridpb}
\Delta^2_B(k)=\frac{k^3}{2\pi^2}\frac{1}{2}
(a\dot{D}/D)^2 L^3 \sum_{\textbf{k}_2} P_{NL}(k_2,z)P_{NL}(|(\textbf{k}-\textbf{k}_2)_p|,z)
\left[W_g(|(\textbf{k}-\textbf{k}_2)_p|)\beta_G(\textbf{k,k}_2)++W_g(k_2)\beta(\textbf{k},(\textbf{k}-\textbf{k}_2)_p)\right]^2.
\end{equation}
In our model, the nonlinearity effect is explicitly quadratic and
larger than that of \citet{Hu00,Ma02}. We will adopt the code of
\citet{Smith02} to calculate $P_{NL}(k,z)$.  

In \S \ref{sec:comparison}, we will test our model against simulations. Our logic is, if the grid version of our model, which captures key
simulation artifacts, gives a good description of the simulation, then, its
asymptotic case, namely the
continuous version of our model would represent an ideal simulation, be free of these simulation artifacts and
describe the real $\Delta^2_B$.

\begin{figure}
\epsfxsize=10cm
\epsffile{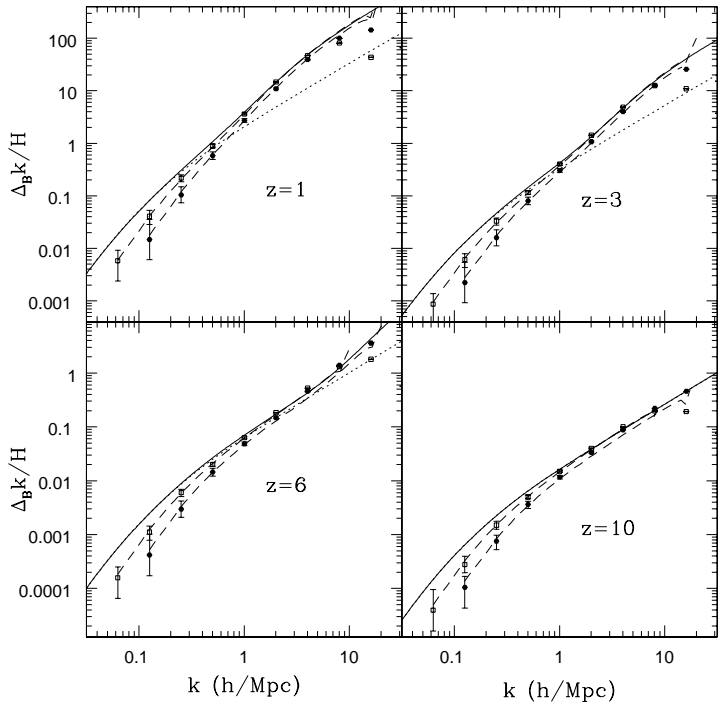}
\caption{The comparison between our model and the simulations with WMAP
cosmology.  The vertical axis is
the dimensionless $\Delta_B k/H(z)$, instead of the $\Delta^2_B$
usually plotted. Two sets of data points are our $1024^3$, $100$ Mpc/h
(open square) 
and $1024^3$, $50$ Mpc/h (filled hexagon) results.  For clarity, we only show a
few data points.  As comparison, we show grid version predictions (dash lines),
continuous version predictions (solid lines) of our model and the the Vishniac
predictions (dot lines). Our grid predictions agree with simulations at all
redshifts and all  scales that simulations can reliably probe.  
\label{fig:pb}}
\end{figure}

\section{Comparison between analytical model and simulations}
\label{sec:comparison}
We first compare the grid version prediction of our model 
Eq. (\ref{eqn:finalgridpb}) with simulations.  Self similar simulations are
ideal to test the effects of box size, 
resolution and non-linearity due to the scaling relation between redshifts and scales (See
 e.g. \citet{Peebles80}). The grid prediction of our model fits simulations very
well at all reliable scales. The cut off due to the finite simulation box size
can 
be seen from the large scale behavior of $\Delta^2_B$  and the resolution
effect can be seen from the small scale behavior of $\Delta^2_B$
(Fig. \ref{fig:self}).  The dependence of the non-linearity effect on the
initial density power index $n$ is naturally explained by our model. The effect
of the non-linearity to the density power spectrum depends on $n$.
For $n=-1$, the non-linearity decreases the non-linear density power spectrum
with respect to the linear density power spectrum.  Our model then predicts a decreasing in $\Delta^2_B$ with respect to the Vishniac prediction by the
non-linearity. For $n=-2$, it is the opposite case. For the WMAP cosmology, the effective power index in
simulation scales is around $-2$ and an amplification of $\Delta^2_B$ by  the
non-linearity is expected, as shown in Fig. \ref{fig:simulation}. 

\begin{figure}
\epsfxsize=10cm
\epsffile{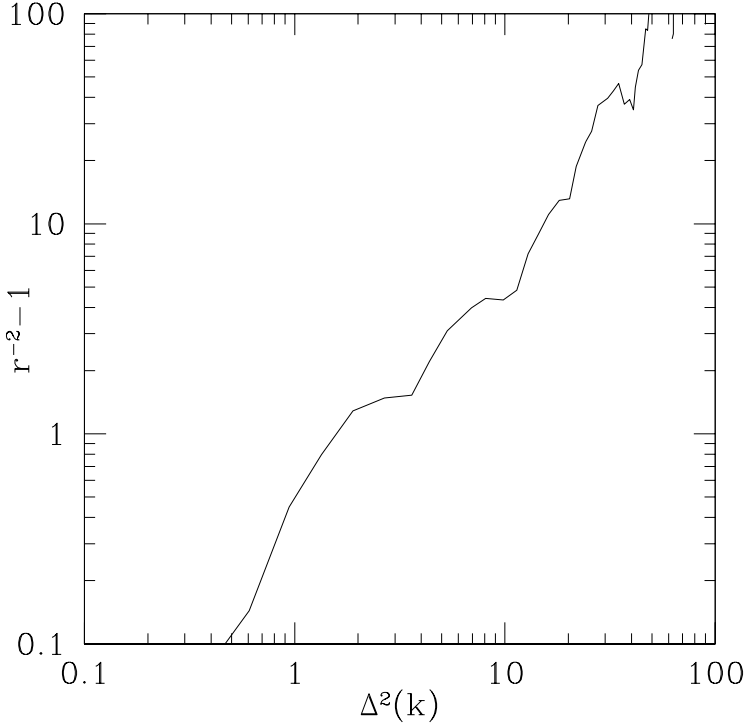}
\caption{The correlation coefficient $r$ between
$\delta$ and $\dot{\delta}$. $r\equiv \Delta^2_{\delta
\dot{\delta}}/\sqrt{\Delta^2_{\delta}\Delta^2_{\dot{\delta}}}$. In the
highly
non-linear regime, $\delta$ and $\dot{\delta}$ are barely correlated
and $r^2$ roughly scales as $1/\Delta_{\delta}^2$. The result shown is for
our WMAP simulation with $50$ Mpc/h box size and $1024^3$ grid
elements. \label{fig:r}}
\end{figure} 

Since our ultimate goal is to provide an accurate description of $\Delta^2_B$
for our real universe, we  test our model in detail against the $1024^3$ simulations with WMAP cosmology.
In order to do that, we need to estimate the variance of $\Delta^2_B(k)$. In
the linear regime $\delta$ is Gaussian distributed and $p_B(\textbf k)\propto \delta \delta
\frac{\textbf k}{k}$, we obtain $\langle p^4_B(\textbf
k)\rangle=\frac{105}{9} \langle p^2_B(\textbf k)\rangle^2$ and
estimate the uncertainty of the simulated $\Delta^2_B$ by this relation.  
Our result (Fig. \ref{fig:pb}) shows that the grid
calculation agrees with simulated $\Delta^2_B$ at all output redshifts and the
full $k$ range that simulations can reliably probe.  Our model passed the test
of all self similar and WMAP simulations we have run and shows a wider range of
applicability than all previous analytical models. Thus we believe that the
grid version of our model is a suitable tool to describe simulated
$\Delta^2_B$. Its continuous version (Eq. \ref{eqn:finalconpb}) then should be
able to describe an ideal simulation 
with infinite box size and resolution and produce the real
$\Delta^2_B$. 

Comparing our analytical prediction and simulations, we concluded that
the 
simulated $\Delta^2_B$ loses power at large scales, gains power at small scales
due to non-linearity until approaching the resolution limit.  Thus it has
a steeper slope. We address several simulation limitations here. (1)  The finite simulation box
size causes the simulated  $\Delta^2_B$ to lose power at large scales. From
Eq. (\ref{eqn:finalgridpb}), $\Delta^2_B$ has contributions from all
scales. The finite box size cuts off the density fluctuation modes with size
larger than half box size. It is these missing modes causing the simulated
$\Delta^2_B$ to lose power at large scales. The smaller the box size, the
more $\Delta^2_B$ power is lost, as can be seen from Fig. \ref{fig:pb}. (2)
The finite resolution causes $\Delta^2_B$ to lose power at less than about
$10$-$20$ grid separation. (3) The 
periodic boundary condition causes the simulated $\Delta^2_B(k)$ to
gain power. This can be seen by the following argument. For the  
cosmology we adopt, $P(k)$ is a decreasing function for $k>0.02$
Mpc/h. Comparing Eq. (\ref{eqn:finalconpb}) and Eq. (\ref{eqn:finalgridpb}),
the periodic boundary condition produces a smaller
$|\textbf{k}-\textbf{k}_2|$ and thus a 
larger $P(|\textbf{k}-\textbf{k}_2|)$ for those
$\textbf{k}-\textbf{k}_2$ which are beyond the simulation $\textbf k$
range.  The case is the 
same for the gas window function and is similar for the kernel
function $\beta(\textbf{k,k}_2)$. But this effect is negligible comparing to
the first two.  

Our modeling of the non-linearity effect works surprisingly well.  For WMAP
cosmology, the
non-linearity does increase $\Delta^2_B$ significantly (fig. \ref{fig:pb}).
As expected, the non-linearity effect
is larger than what \citet{Hu00,Ma02} predicted since their models omit the
contribution of ${\bf v}_B$.  Our model passes the test of all of our
simulations and suggests that it not only captures the contribution of ${\bf
v}_B$ qualitatively but also quantitatively. Though it is hard to explain why
it works well quantitatively, we believe that it gives a reasonable
description of $\Delta^2_B$ in the non-linear regime.  

Our model could be understood from the equi-partition of $\bf{p}$ in
the non-linear 
regime, which states $\Delta^2_B=2\Delta^2_E$, as found in our
simulations, and the mass conservation equation, which states
$\dot{\delta}+\nabla \cdot {\bf p}_E/a=0$. Defining the cross
correlation coefficient $r\equiv \Delta^2_{\delta
\dot{\delta}}/\sqrt{\Delta^2\Delta^2_{\dot{\delta}}}$ between
$\delta$ and $\dot{\delta}$, one then has
\begin{equation}
\Delta^2_B=\frac{1}{2} \frac{a^2}{k^2}
\frac{\dot{\Delta^2(k)}^2}{r^2\Delta^2(k)}.
\end{equation}
$r$ in the linear theory and the stable clustering regime can be
calculated analytically. But since possibly due to the ongoing merger
process, the stable 
clustering may not hold even in highly non-linear regime
\citep{Smith02}, for  most non-linear regimes, $r$ is hard to
estimate, even in an interpolation way. In our simulations, we find
that in highly non-linear regime, $\delta$ and $\dot{\delta}$ becomes
weakly correlated . $r^2$ roughly scales as $1/\Delta^2$
(Fig. \ref{fig:r}). This scaling gives $\Delta^2_B\propto \Delta^4$,
which is consistent with our previous model (Eq. \ref{eqn:finalconpb}
and \ref{eqn:finalgridpb}).

Simulation limitations put strong requirements on the simulation
power. For example, from Eq. (\ref{eqn:finalgridpb}), we estimate that, in order to produce a reliable $\Delta^2_B$ at $ 0.1$ h/Mpc
$<k<10$ h/Mpc, a hydro simulation with a box size $L\geq 400$ Mpc/h and
resolution $N>4096$ is required, which is not realistic at present. But the
continuous version of our
analytical model (Eq. \ref{eqn:finalconpb}) is free of these simulation
limitations and is able to produce a reliable $\Delta^2_B$ at all $k$
and $z$ range relevant to the kinetic SZ effect.  Hereafter, we will adopt
Eq. (\ref{eqn:finalconpb}) to calculate the kinetic SZ power spectrum.

\section{The kinetic Sunyaev Zel'dovich effect}
\label{sec:cl}
\begin{figure}
\epsfxsize=10cm
\epsffile{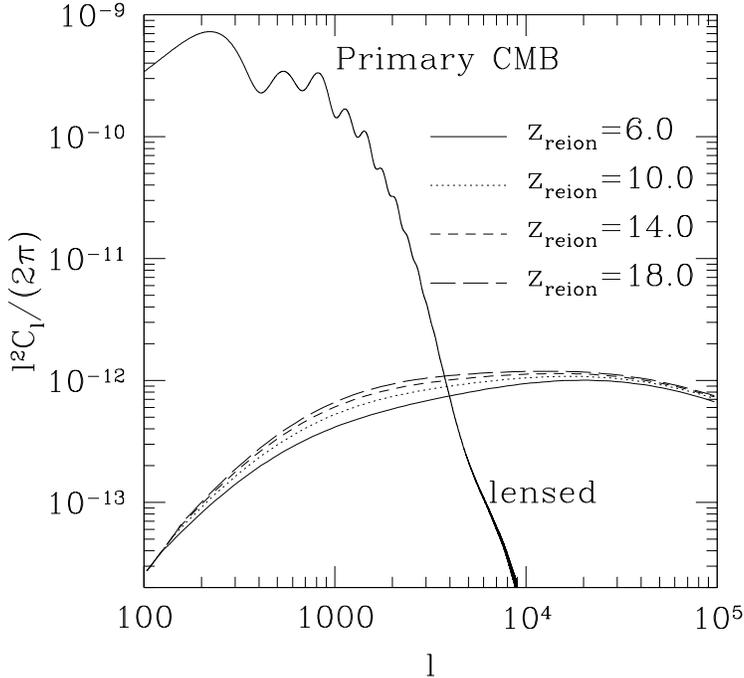}
\caption{The kinetic SZ $C_l$ dependence on scales and reionization
redshift. The primary CMB is the lensed CMB calculated by CMBFAST
\citep{CMBFAST}. For
$6<z_{\rm reion}<20$ and $3000<l<9000$,  $l^2C_l/(2\pi)\simeq 0.91
\times 10^{-12} 
[(1+z_{\rm reion})/10]^{0.34}(l/5000)^{0.23-0.015(z_{\rm
reion}-9)}$. \label{fig:kszclz}}

\end{figure}
\begin{figure}
\epsfxsize=10cm
\epsffile{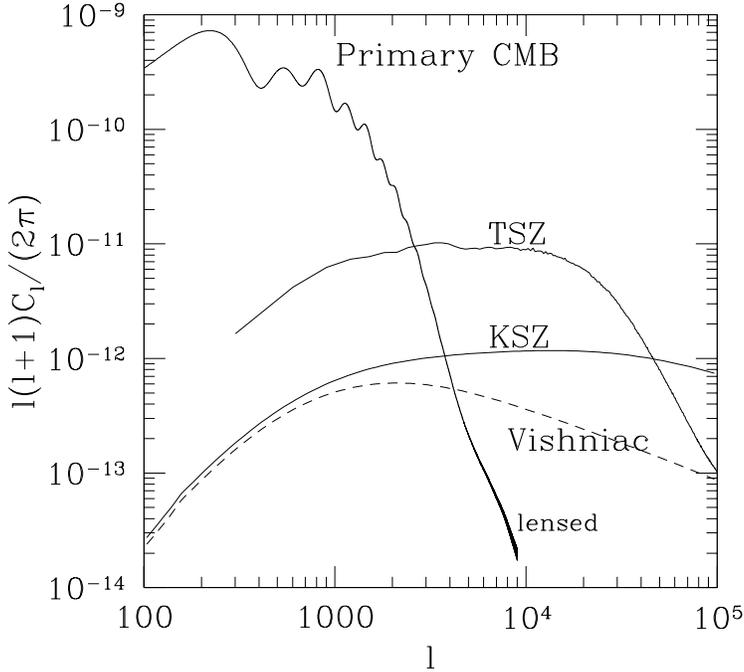}
\caption{The kinetic SZ
  effect power spectrum (solid line). The WMAP cosmology with $\Omega_m=0.268$,
$\Omega_{\Lambda}=1-\Omega_0$, $\Omega_b=0.04$, $h=0.71$, $\sigma_8=0.84$ is
  adopted. We assume that the universe suddenly ionized at $z=16.5$ and remain
completely ionized after that. To show a comparison, we show the primary CMB
power spectrum and the Thermal SZ power spectrum, which is adopted from a $512^3$ MMH
simulation \citet{Zhang02} and has been scaled to the WMAP cosmology by the
scaling relation $C_l \propto \sigma_8^{\sim 7}$. The dash line is the Vishniac
prediction. The non-linearity increases the kinetic SZ power spectrum by a
factor of $2$ at $l\sim 4000$. \label{fig:cl}}
\end{figure}

The kinetic SZ power spectrum $C_l$ is very sensitive to  the reionization
history. To see this, we estimate its 
dependence on the reionization redshift. For simplicity, we assume that the
universe 
is reionized at once at $z_{\rm reion}$ and remains completely ionized after
that. Since $\tau\ll 1$, we omit its effect temporarily for the
estimation. First, we consider the linear case. Combining Eq. (\ref{eqn:cl}) \& (\ref{eqn:conpb}), we obtain
\begin{eqnarray}
C_l(z_{\rm reion}) &\propto& \int^{z_{\rm reion}} (1+z)^4 (a\dot{D}D)^2 (\chi(z))^{1-n_B} \frac{{\rm
d}\chi}{{\rm d}z} dz \propto \int^{z_{\rm reion}} (\frac{d D}{d a} D)^2
\sqrt{(\Omega_0(1+z)^3+\Omega_v)}(\chi(z))^{1-n_B}  dz \\ \nonumber
&\propto& \int^{z_{\rm reion}} (1+z)^{(1-n_B)\alpha-0.5} dz \propto (1+z_{\rm
reion})^{(1-n_B)\alpha+0.5}\ \ \ \ \ \ {\rm for}\ z\gg 1
\end{eqnarray}
Here, $n_B$ is the power index of $\Delta^2_B(k,z)$ at $k=l/\chi(z)$. For
$l\geq 10^3$ and $z>1$, $n_B\la 1$. We approximate $\chi(z)\propto
(1+z)^{\alpha}$. For $z\geq 2$, $0<\alpha<0.5$. For $z\geq 10$, $0<\alpha<0.1$.
When $z\geq 2$, $D\propto a$ and $\frac{dD}{da}\propto 1$.
Then, for $z>2$, $C_l$ diverges toward high
$z_{\rm reion}$ with a scaling relation $C_l\propto (1+z_{\rm reion})^{\sim
0.5}$.  The non-linearity increases the contribution from low redshifts and
makes the $C_l$ redshift dependence weaker.  The damping caused by the
optical depth $\tau$ causes further suppression. 
We show $C_l$ calculated for various $z_{\rm reion}$
by Eq. (\ref{eqn:cl}) \& (\ref{eqn:finalconpb})  in
Fig. \ref{fig:kszclz}. The $C_l$ behavior has three distinct regions. (1)
$l\ll 1000$. $C_l$ in this region is mainly contributed by low redshift linear
regions. Its dependence on $z_{\rm reion}$ is weak. (2) $l\gg 10000$. $C_l$
in this region is mainly contributed by low redshift highly non-linear
regions. Its dependence on $z_{\rm reion}$ is also weak. (3) $1000\la l\la 10000$. $C_l$ in this region has large contributions from the high redshift
universe and thus has the highest  sensitivity to $z_{\rm reion}$. As we will
discuss in \S \ref{sec:ACT}, this region is also observationally accessible
in the
near future. For $6<z_{\rm  reion}<20$ and $3000<l<9000$, our result can be fitted by
\begin{equation}
l^2C_l/(2\pi)\simeq 0.91 \times 10^{-12}
[(1+z_{\rm reion})/10]^{0.34}(l/5000)^{0.23-0.015(z_{\rm reion}-9)}.
\end{equation}
The error introduced by this fitting formula is less than $2\%$. The
corresponding  temperature fluctuation $\Delta T$ is then
\begin{equation}
\Delta T_l\simeq 2.60 
[(1+z_{\rm reion})/10]^{0.17}(l/5000)^{0.115-0.0075(z_{\rm reion}-9)} \mu {\rm
K}.
\end{equation}

For the WMAP favored  $z_{\rm
reion}=16.5$, we show $C_l$ predicted by  our model and by the corresponding 
Vishniac effect in fig \ref{fig:cl}. The non-linearity increases the power
spectrum at small angular scales. At $l\sim 4000$, this amplification reaches
$2$. The non-linearity then produces a broad peak in the power spectrum, which 
extends from $l\sim 3000$ to 
$l\sim 10^5$ with an amplitude $\Delta T\simeq 2.7 \mu $K.  The large $l$ behavior will be certainly affected by the gas window
function  $W_g(k)$. Its effect can be easily incorporated, as discussed in \S \ref{subsec:linear}.  Since it mainly
affects the $l>10^4$ region, where it is hard to observe, we will not discuss
its effect here.

\section{The power for the kinetic SZ effect  to constrain the reionization history}
\label{sec:ACT}

\begin{figure}
\epsfxsize=10cm
\epsffile{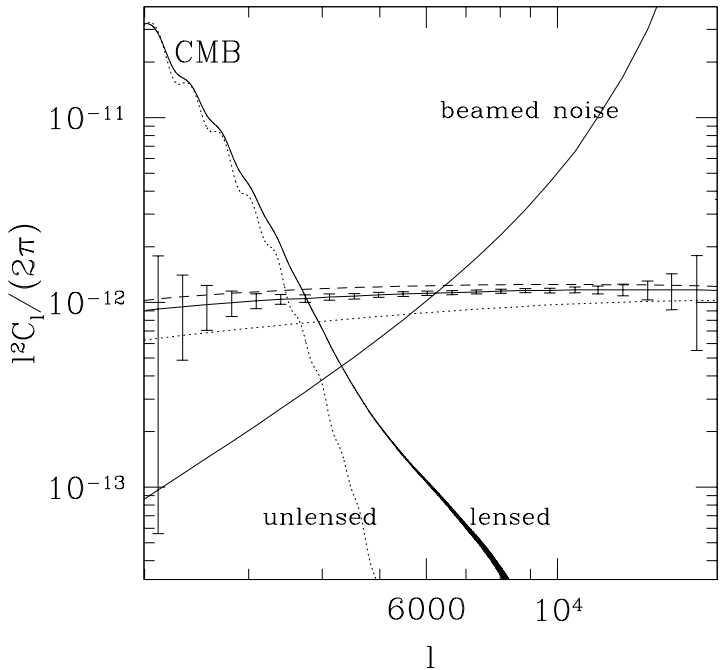}
\caption{The sensitivity of ACT to distinguish reionization scenarios. The three
unlabeled lines are reionization scenarios A, B and C, from bottom up. The
error bars are the estimated ACT errors for the reionization scenario B.  At
$l\sim 6000$, the error is $\sim 3\%$. B and C are degenerate in the
primary CMB measurement but they can be distinguished by ACT with $>3\sigma$
confidence. One can  bin the data points at $3000<l<8000$ and do not lose
useful information since the power spectrum is almost flat at this range, thus
one can reach $1\%$ accuracy in the $C_l$ measurement and B and C can be
distinguished by $10\sigma$ confidence. For the single reionization scenario, the
reionization redshift can be constrained with better than $3\%$ accuracy. \label{fig:ACT}}
\end{figure}

For $z_{\rm reion}\sim 7$, as  favored by the Lyman-$\alpha$
absorption observation versus  $z_{\rm reion}\sim 17$, as
inferred from WMAP, $C_l$ would differ by $45\%$. This sensitive 
dependence of $C_l$ on the reionization history opens a window for future CMB
observations to constrain the reionization history and break the degeneracy met
in the primary CMB argument. 
The power of the kinetic SZ effect to constrain the reionization history
depends on the accuracy of the kinetic SZ effect measurement. In this section,
we take the Atacama cosmology telescope (ACT) as our target to
address this issue. 

ACT will operate at three bands, $145$ Ghz, $225$ Ghz and $265$ Ghz,
measure over $100$ square 
degree sky with  noise error $\sigma_N \simeq 2 \mu$K per $1.7^{'}\times1.7^{'}$
pixel. At the $225$ Ghz band, the thermal SZ signal is negligible and we take
this band to estimate the sensitivity of ACT to measure the kinetic SZ effect. The noise power spectrum is given by $C_N=4\pi f_{\rm pix}\sigma_N^2$. Here,
$f_{\rm pix}$ is the fractional sky coverage of each pixel.  The error in the kinetic SZ power spectrum measurement comes
from three sources: instrumental noise, primary CMB and cosmic variance. If the
primary CMB power spectrum and the noise spectrum are known and can be
deducted, we
have then 
\begin{equation}
\frac{\Delta
C_l}{C_l}=\sqrt{\frac{(\eta+2)+2(C_N/C_l/W_N^2)^2+2(C_{CMB}/C_l)^2}{(2l+1)\Delta
l 
f_{\rm sky}}}. 
\end{equation}
$W_N(l)$ is the Fourier transform of the natural beam function. For the $225$
Ghz band, the ACT resolution is $1.1^{'}$. We then 
approximate the beam function as  a top-hat function with an effective radius
$\sqrt{1.1^{'}\times 1.1^{'}/\pi}$. We normalize it to have $W_N(l\rightarrow 1)\rightarrow 1$. $f_{\rm
sky}$ is the fractional sky coverage of the whole survey area, which we adopt
as $100$ square degree. $\eta$ is a measure of the
non-Gaussianity, which will be taken as $105/9-3=26/3$, as in the 3D case obtained from the
second order perturbation theory (\S \ref{subsec:linear}). Due to the
projection, the 2D kinetic SZ non-Gaussianity would be weaker and the error
obtained above should serve as an upper limit.

We consider three reionization scenarios: (A) the universe reionized once at
$z_{\rm reion}=7$, (B) the universe reionized completely
once at $z_{\rm reion}=16.5$, (C) the universe reionized completely at $z=21$
and became neutral at $z=13$ and reionized again at $z=7$. The last two
has roughly the same optical depth 
$\tau=0.174$ to the last  scattering surface and are degenerate in the primary
CMB measurement. In Fig. \ref{fig:ACT}, we forecast the accuracy for  ACT to
measure the  kinetic SZ $C_l$ assuming the reionization scenario B. $\Delta
l=l/10$ is adopted. At $l\la 2500$, primary CMB dominates and prohibits the extraction of the
kinetic SZ signal.  At $l\ga 10^4$, the instrumental noise
dominates.  At $l\sim 6000$, ACT will reach its highest sensitivity with a
$3\%$ accuracy in the kinetic SZ $C_l$ measurement. Since the kinetic SZ power spectrum is almost flat at $l>3000$, one
can bin the data points ($\sim 10$ points, fig. \ref{fig:ACT}) at $3000<l<8000$
and do not lose any crucial
information. Binning 
in this way, the ACT-measured $C_l$ accuracy will reach better than
$1\%$.

The accuracy of inferred reionization history depends on both the
precision of KSZ power spectrum measurement and theoretical
predictions.  Given perfect theoretical predictions, for the single
reionization scenario,  the
uncertainty in the reionization redshift $\Delta z_{\rm reion}/z_{\rm reion}
\simeq\Delta C_l/C_l/0.34 \sim 3\%$.  The two CMB-degenerate
reionization scenario B \& C can be distinguished with $>10\sigma$
confidence\footnote{We assume different bins are uncorrelated. This assumption is exact
for a whole sky survey. For a survey with limited sky coverage, this
assumption only roughly holds in our case. For two successive data points of Fig. \ref{fig:ACT} at
$3000<l<8000$, the corresponding spacial 
separation at $z\sim 10$ is $\sim 1$ Mpc/h, which is about the same as  the
density correlation length at $z=10$. So, one can safely neglect cross
correlations in different bins.}. But uncertainties in the KSZ signal
extraction and theoretical prediction may degrade the power of ACT to measure
the kinetic SZ effect. 

One is the lensing effect. Since lensing changes CMB photon positions
but not  CMB temperature, it results in a smoothing of the CMB power
spectrum. Because the primary CMB power spectrum drops quickly after
$l\sim 2000$,   lensing amplifies it by large factors at small angular
scales (Fig. \ref{fig:ACT}).  
In our estimation, we have assumed that this lensed CMB power spectrum can be
predicted and subtracted. Since the lensing effect  at these scales involves
non-linear dark matter clustering, the precise prediction of its power spectrum
could be difficult.  If we take this possibility, the residual CMB (lensed
subtracted unlensed) would prohibit the measurement of KSZ at $l<4000$ and
increase the error of KSZ $C_l$ measurement at $l\in[4000,6000]$ to be larger
than $10\%$. But its effect at $l>6000$ is small. Binning the $\sim 4$ data points at $7000<l<9000$ would still
give a better than $2\%$ accuracy. Furthermore, the future lensing survey such
as the \citet{CFHT} will measure
the dark matter distribution to $1\%$ accuracy and allow a precise
prediction of the lensed CMB power spectrum. So, the lensing effect of the
primary CMB could be well handled. Lensing also changes the KSZ power
spectrum. But since the KSZ power spectrum
is almost flat, this lensing effect is negligible. 

In our discussion, we have assumed the thermal SZ effect to be negligible at
$\sim 225$ Ghz. This is true in the non relativistic limit. The
relativistic correction (see e.g. \citet{Dolgov01}) produces a 
temperature decrement at this band and its effect on ACT measurement
of cluster SZ effect has been discussed
\citep{Aghanim02}. This relativistic correction is of the order $k_B T_e/(m_e
c^2)\simeq T_e/500$keV. Though it is non-trivial for the measurement
of high temperature 
cluster SZ effect, 
it is negligible for a blank sky SZ power spectrum measurement. The mean
electron temperature at present due to gravitational heating is expected to be less than $0.5$ keV 
\citep{Pen99,Zhang01}. COBE/FIRAS puts an upper limit $1.5\times 10^{-5}$
\citep{Fixsen96} on the
mean 'y' parameter and basically excludes dramatic feedback with $T\ga
1$ 
keV.  One then expects the mean electron temperature at present to be
$\la 1$ keV. Since the blank sky thermal SZ power spectrum at $l>4000$ is mainly contributed
by these low temperature IGM gas at $z\ga 0.5$ \citep{Zhang01}, we expect
that the relativistic correction is much less than $1\%$ and is thus
negligible. 

We have not considered the contamination of unresolved IR and radio
sources, whose effects are estimated to be non-negligible
\citep{White03}.  At small angular scales, these sources are present as
Poisson  noise in the  
KSZ measurement. From the large $l$ power spectrum behavior, the power
spectrum of the sum of instrumental noise and this 
Poisson noise can in principle be measured and
subtracted. Multi-frequency observations also help to clean them out. 

We estimate the dependence of the KSZ power spectrum on cosmological
parameters. One can
easily work out that $\Delta^2_B(k,z)\propto
\beta^2H^2(z)\sigma_8^{4\sim 6}$ ($4$ in the linear regime and $6$ in
the highly nonlinear regime where stable clustering holds). Here,
$\beta\equiv adD/da/D\simeq 
\Omega_m^{0.6}(z)$.   In our interested $3000\la l\la 10000$, $C_l$ contributions
mainly come from $z>1$, where $\beta\simeq 1$, $H(z) \propto
\Omega_0^{1/2}$ and the comoving distance $x$ is roughly $\propto
\Omega_0^{-\gamma}$ with 
$0<\gamma<1/2$. Here, $\Omega_0\equiv \Omega_m(z=0)$. Applying the above relation to Eq. \ref{eqn:cl}, we
find that   $C_l \propto (\Omega_b
h)^2 \Omega_0^{1/2+\gamma(n_B-1)}\sigma_8^{4\sim 6}\propto (\Omega_b
h)^2 \Omega_0^{\la 1/2}\sigma_8^{4\sim 6}$. As a reminder, $n_B$ is the power
index of $\Delta^2_B(k,z)$ at $k=l/x$ and $0<n_B\la 1$ for $l\ga 10^3
$ and $z\ga 1$. So there is a degeneracy between $z_{\rm reion}$,
$\Omega_b h$ and $\sigma_8$ and the recovered $z_{\rm reion} \propto
(\Omega_b h)^6 \sigma_8^{\sim 15}$. Thus, current uncertainties 
in cosmological parameters degrade the
accuracy of the extracted $z_{\rm reion}$ significantly. But this
degenerancy is not that bad as appears since it is the
matter density power spectrum $P_{NL}(k)$ at Mpc/h scale, instead of
$\sigma_8$,  that directly  determines the kinetic SZ effect at $l\sim $several
thousands through $C_l \propto P^2_{NL}$.  $P_{NL}(k)$ at Mpc/h scale 
will be measured by CFHT  
legacy survey with $1\%$ accuracy (see \citet{Pen03} for the
discussion of relevant length scales in the recovered 3D power
spectrum). Future
experiments such as Planck, SNAP and CFHT would significantly reduce the
above uncertainties in cosmological parameters and density power
spectrum \citep{Tegmark02} and allow a precise recovery of the
reionization history   
from the kinetic SZ effect.

At the presence of various noise sources in KSZ
measurement discussed above, ACT would still be able to  measure
$C_l$ with $\sim 2\%$ accuracy. Given a sufficient understanding of
cosmic baryonic matter density and 
the matter density power spectrum, which will be measured precisely by
future experiments, the reionization redshift could be constrained
with several percent accuracy, some multiple reionization scenarios
degenerated in the primary CMB temperature and T-E polarization
measurement could be distinguished by $\sim 10\sigma$ confidence.

By far we only discussed the uniformly reionized universe. Patchy
reionization introduces extra power to the KSZ effect and may degrade the
power of KSZ to constrain the reionization history. But future
observations may be able to separate it by its characteristic
$C_l$ behavior. Furthermore, by cross correlating the KSZ effect with
the cosmic 21 
cm background, one can further recover the redshift information of
the KSZ effect at high redshifts and extract the contribution from the patchy
reionization. Since in the 
literature,  the significance of  patchy reionization is  under debate
(e.g. \citet{Gnedin01,Valageas01,Santos03}), we postpone its further discussion.

\section{Summary}
The dependence of
the kinetic SZ effect on the reionization history allows a detailed probe of
the reionization process and the feasibility to break the degeneracy posed in
the primary CMB measurement. This relies on the precision of the kinetic
SZ effect measurement, which will be realized by future CMB experiments such as
ACT and precision modeling of the kinetic SZ effect. We have built an
analytical model 
for the kinetic SZ effect and tested against our high resolution hydro
simulations through the 3D power spectrum $\Delta^2_B(k,z)$ of the
momentum curl part, which completely determines the kinetic SZ effect
angular power spectrum. The grid  version of our model, which captures various
limitations of simulations such as the finite 
box size effect and  thus reproduces simulation results at all available
simulations scales
and redshifts with better than several percent accuracy. The continuous version, which represents an
ideal simulation with infinite box size and resolution and is free of
simulation limitations, should describe the real $\Delta^2_B$ to a high
accuracy.   
Comparing between simulations and analytical predictions, we found that 
\begin{itemize}
\item The simulated $\Delta^2_B$ loses power at large scales due to
the cutoff of the density fluctuation beyond the simulation box. This effect
causes the simulated $\Delta^2_B$ to be steeper than the real one at large
$k$. The smaller the simulation box, the severe the problem is.
\item The simulated $\Delta^2_B$ loses power at small scales due to the finite
resolution.
\item Correspondingly, the simulated kinetic SZ power spectrum loses power at
both large and small scales. In order for simulation to directly
produce reliable
$C_l$ at $1000<l<10000$ without reference to our models, 
a hydro simulation with box size $>400$ Mpc/h and
resolution $>4096^3$ is required.

\item The effect of  the non-linearity on $\Delta^2_B$ has the same tendency
as its effect on the density power spectrum. Its effect can be modeled by substituting both linear
density power spectra in the Vishniac integral (Eq. \ref{eqn:conpb} \& \ref{eqn:gridpb}) with the
corresponding non-linear ones (Eq. \ref{eqn:finalconpb} \&
\ref{eqn:finalgridpb}). For WMAP cosmology, the non-linearity increases the
density power spectrum and $\Delta^2_B$.

\item Our analytical model in the grid version captures all main features of
simulated $\Delta^2_B$ and reproduces the simulation results with high accuracy. Its continuous
version is an accurate description of the real $\Delta^2_B$ and therefore the
KSZ effect over a wider parameter range in power spectrum slope than the
past models proposed in the literature.

\item For the WMAP cosmology and the single reionization scenario, the kinetic
SZ power spectrum can be approximated with better than several percent accuracy
as $l^2C_l/(2\pi)\simeq 0.91 \times 10^{-12}
[(1+z_{\rm reion})/10]^{0.34}(l/5000)^{0.23-0.015(z_{\rm reion}-9)}$ for
$6<z_{\rm reion}<20$ and $3000<l<9000$. This
strong reionization redshift dependence will allow a precision measurement of
the reionization history. For the
WMAP favored $z_{\rm reion}\simeq 16.5$, the kinetic SZ power spectrum has a broad peak extending from $l\sim 3000$-$10^5$ with an
amplitude $\simeq 2.7\mu$k.   Among cosmological parameters,
$\sigma_8$ and $\Omega_b$ are the two that $C_l$ is most sensitive to.
$C_l$ scales as $(\Omega_b h)^2 \sigma_8^{4\sim 6}$.
\item  ACT will measure the kinetic SZ effect
to $1\%$ accuracy. Given precise measurements of cosmic baryonic
density $\Omega_b h$ and matter density power spectrum,  for the  single
reionization scenario, the reionization redshift $z_{\rm reion}$ can be
constrained with several percent accuracy. The kinetic SZ effect can
further  distinguish some more complicated reionization scenarios and
breaks the degeneracy met in the primary CMB measurement with $\sim
10\sigma$ confidence.
\end{itemize}
\section*{Acknowledgments}
We would like to thank Uros Seljak and Steen Hansen for helpful discussions. We
thank Lyman Page for providing information about ACT. We thank Asantha
Cooray, Lloyd Knox and Mario Santos for the discussion of patchy
reionization. 
P.J. Zhang thanks the department of Astronomy \& Astrophysics and CITA
at the University of Toronto where part of
the work was done. P.J. Zhang is  
supported by the DOE and the NASA grant NAG 5-10842 at Fermilab.
Computations were performed on the CITA Pscinet computers funded by
the Canada Foundation for Innovation.

\end{document}